\documentclass[fleqn,usenatbib,onecolumn]{myarticle}

\usepackage[T1]{fontenc}
\usepackage{ae,aecompl}

\usepackage{graphicx}	
\usepackage{amsmath}	
\usepackage{amssymb}	

\usepackage{mathabx} 
\usepackage{graphicx}
\usepackage{hyperref} 
\usepackage{subfigure}  
\usepackage{color}
\usepackage{ulem} 
\usepackage{lineno}

\title[Time-Stable Solar Periodicities]{Stable Solar Periodicities: The Time Stability \footnote{Published in Monthly Notices of the Royal Astronomical Society, which is available at \url{https://doi.org/ }}}

\author[K. Chol-jun et al.]{
Kim Chol-jun,$^{1}$\thanks{E-mail: cj.kim@ryongnamsan.edu.kp}
Jon Kyong-phyong$^{1}$
\\
$^{1}$Faculty of Physics, \textbf{Kim Il Sung} University, DPR Korea
}

\date{\today} 

\pubyear{}

\begin{document}
\label{firstpage}
\pagerange{\pageref{firstpage}--\pageref{lastpage}}
\maketitle

\begin{abstract}
We show that while the Fourier transform can figure only an ``intensity'' of a periodic signal, there is an additional information embedded, which is a ``coherence'' of the signal. Supposing that the periodicity is reflected only on the ``coherence,'' we introduce a time stability as a measure of ``coherence'' excluding the ``intensity,'' i.e. amplitude of signal. In stead of classical strength-based significance, where strength implies the power or amplitude, we adopt a stability-based significance as criterion to choose cycles in a random signal. We inspect the time-stable solar periodicities and show that most periodicities discovered in exterior solar activities such as the solar wind inhere in interior solar activity such as the sunspot and that the time stability can be an effective tool in spectral analysis of stochastic solar activity.
\end{abstract}


\section{Introduction}\label{sec:intro}

The search for the solar periodicities has a long history. After Schwabe discovered the 11-yr cycle, Hale found the 22-yr cycle of the solar magnetism and \cite{Gleissberg1939} indicated a long-term variation of $80\sim90$ yr in the 11-yr cycle amplitudes . 

However, a great randomness of solar activity makes it difficult to find any other periodicities. For example, \cite{Kane1977} indicated many periods of 0.7, 0.8, 1.0, 1.3, 1.5, 1.75, 2.0, 2.3, 2.6, 2.9, 3.3, 3.6, 4.1, 4.7, 5.2, 6.0, 7.0, 8.5, 10.0, 12.0, 15.8, 20.0, 26.0, 32.0, 40.0, 50.0, 66.0 and 80.0 yr with an analysis of superposition of time-lagged sequences. Even in the Fourier analysis and the maximum entropy method (MEM) approach there appear innumerable periodicities \cite[for example, see][]{Stuiver1989}. We need a criterion to choose meaningful periodicities among them, and we usually select periodicities of highest or significant peak.

Even now periodicities are being reported in various solar phenomena with new methods of spectral analysis: the 27-day period (the Dicke cycle) of solar extreme UV irradiance \citep{Vita-Finzi2010}, the 152-156 day periodicity in the solar flare occurrence rate and sunspot areas \citep{Rieger1984, Bai1987, Bai1990, Lean1990, Krivova2002}, the periodicity of 1.3 yr observed intermittently in solar wind speed and geomagnetic index \citep{Richardson1994, Paularena1995, Krivova2002}, the quasi-biennial oscillation in solar magnetism and geomagnetism \citep{Benevolenskaya1995, Mursula2003, Berggren2009, Hathaway2015}, an intermittent 2.5-yr variation in the solar neutrino flux \citep{Shirai2004}, a peak of 14.5 yr in spectrum of the monthly sunspot number series \citep{Petrovay2010a, Petrovay2010b}, the three-cycle quasi-periodicity (TRC) in solar and geophysical data \citep{Ahluwalia1998, Du2006}, the 5-cycle periodicity in the maximum amplitudes of the modern era sunspot cycles \citep{Du2006}, a 44-66 yr periodicity in the North-South asymmetry of the minimum and a 130-140 yr periodicity in the asymmetry of the maximum \citep{Javaraiah2019} and other periodicities \citep[for example, see][]{Nagovitsyn2004, Berggren2009}.

The reconstruction of the past solar activity based on the cosmogenic radionuclides such as $^{14}$C and $^{10}$Be \citep{Solanki2004, Steinhilber2012, Wu2018} allows us to analyze the long-term solar periodicities: the Gleissberg (88-yr), the Suess/de Vries (208-yr) and the Hallstatt cycles ($\sim$2400-yr) show significant appearances in most analyses \citep{Usoskin2016, Beer2018, Chol-jun2020} and the Eddy ($\sim$1000-yr) and unnamed $\sim$350-yr as well as $\sim$500 and $\sim$710-yr periodicities have been reported, though less significant \citep{Usoskin2008}. 

The periodicities have been found in the historical records: the Suess/de Vries and the Gleissberg periodicities in the historical naked-eye sunspot records \citep{Xu1990, Ogurtsov2002, Ma2009b, Chol-jun2020} and the Gleissberg and Attolini's (130-yr) cycles in the records of historical aurorae \citep{Attolini1988}. Although the historical records have uneven spacing and vague magnitude and the reported periodicities have wide variance, the periodicities inferred from the strongly noisy records should have a robustness.

In this paper we use a new method based on a stability of cycle in stead of the traditional methods based on strength of cycle such as power or amplitude and try to find periodicities consistent among different sunspot-related datasets of various time scales:
the daily and monthly sunspot number (DSSN \& MSSN), the yearly international SN series (SILSO ISN version 2 - YSSN), the daily sunspot area (DSSA), the daily, monthly and yearly Group Sunspot Number (GSN) series (DGSN, MGSN \& YGSN), 10-yr-spanned reconstructed sunspot number (RSSN) and 22-yr-spanned reconstruction of total solar irradiance (RTSI).
\footnote{\label{footnt:SSN} The daily (1818/01/01 to 2013/12/31) and monthly (1749/01 to 2015/05) SSN datasets are available at \url{http://www.ngdc.noaa.gov/stp/space-weather/solar-data/solar-indices/sunspot-numbers/international/tables/}. Because of some missed early records and length-consistency between daily datasets or monthly datasets, the daily dataset is taken here only from 1899/1/4 to 2013/12/31 and the monthly from 1815/01 to 2014/12. 

The SILSO ISN version 2 dataset (1700.5 to 2018.5) is available at \url{http://www.sidc.be/silso/DATA/SN_y_tot_V2.0.txt}. See also \citet{Clette2014}.  

The daily sunspot area dataset (1874/05/01 to 2016/10/31) is available at \url{https://solarscience.msfc.nasa.gov/greenwch/daily_area.txt}. Because of the aforementioned reason, the dataset is taken here only from 1901/11/5 to 2016/10/31.

The yearly (1610.5 to 1995.5), the monthly (1610/01 to 1995/12) and the daily (1610/01/01 to 1995/12/31) GSN datasets are available at \url{http://www.sidc.be/silso/groupnumberv3}. See also \citet{Hoyt1998}. Because of the aforementioned reason, the daily dataset is taken here only from 1881/1/3 to 1995/12/31 and the monthly from 1796/1 to 1995/12. 

The length of all daily datasets is 42000 and that of monthly datasets is 2400. The length consistency is useful to avoid repetition of long calculation especially in Monte Carlo simulation for statistics.

The RSSN dataset is available through the MPS sun-climate web-page at \url{https://www2.mps.mpg.de/projects/sun-climate/data/SN_composite.txt}. The dataset covers the period from 6755 B.C. to A.D. 1885 by decadal interval. See also \citet{Wu2018}.

The RTSI dataset is available through the NOAA web-page at \url{https://www1.ncdc.noaa.gov/pub/data/paleo/climate_forcing/solar_variability/steinhilber2012.txt}. The dataset covers the period from 7439 B.C. to A.D. 1977 by 22-yr interval. See also \citet{Steinhilber2012}.}

\section{Risks of the traditional spectral analysis}\label{sec:risk}
There appear usually innumerable peaks in spectrum of a random signal. We use the strength-based significance to distinguish intrinsic or meaningful cycles among those numerous noisy peaks, where strength implies the power or amplitude of cycle. In this way, we consider only significant or highest peaks and neglect low peaks. 

\begin{figure*}
\centering
\subfigure[]{\label{fig:SigniPerWave} \includegraphics[width=0.65\textwidth]{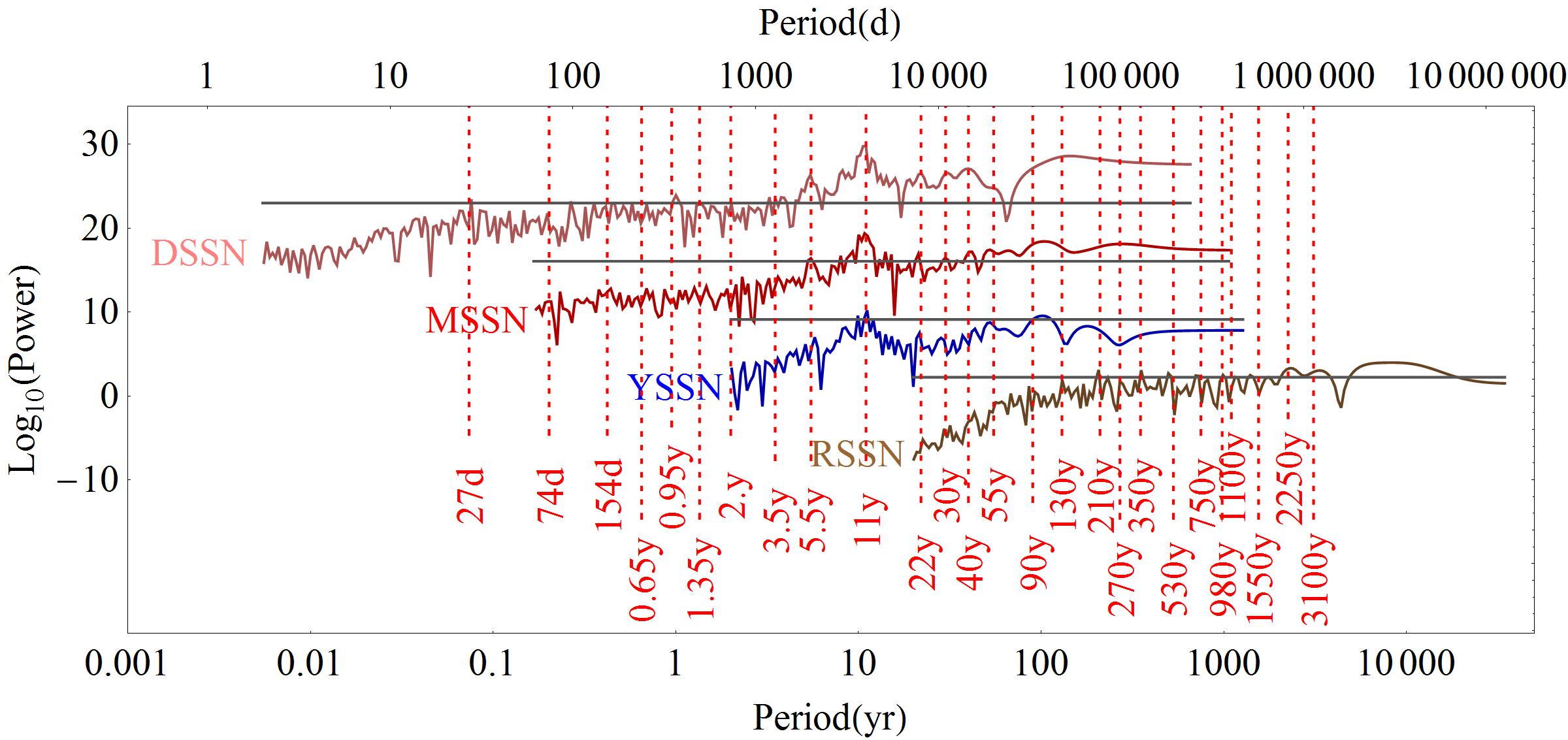}}
\subfigure[]{\raisebox{2.0ex}[0ex][0ex]{\label{fig:Peakform} \includegraphics[width=0.2\textwidth]{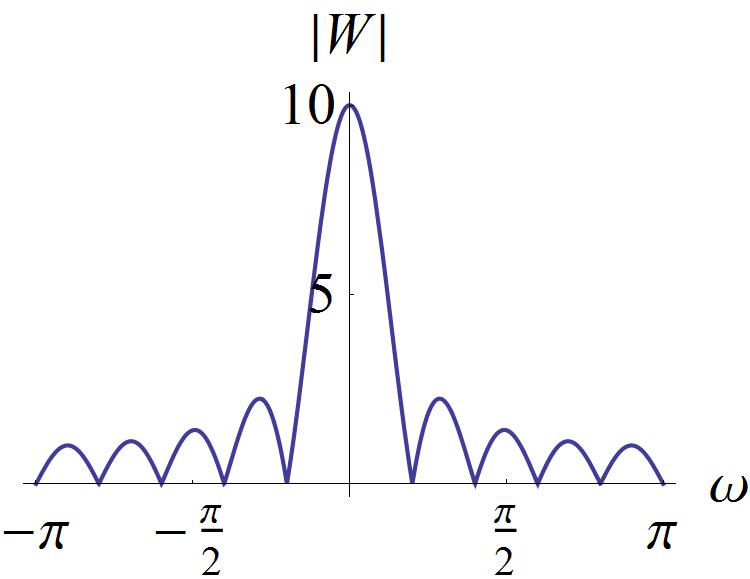}}}
 \caption{ (a) The Lomb-Scargle periodogram for the sunspot-related datasets. The horizontal lines stand for the FAP of 0.01 for each datasets. The spectra of DSSN, MSSN and YSSN are shifted vertically by $10^9$, $10^6$ and $10^3$ times, respectively. The periodicities are denoted only for reference so they do not correspond necessarily to the peak. (b) $\vert W\vert$ vs. $\omega$ in Eq.~\ref{eq:W}. This also corresponds to the shape of peak in spectrum for a mono-cyclic time series of unit amplitude.}
\end{figure*}

A typical example is the Lomb-Scargle periodogram \citep{Lomb1976, Scargle1989}. The Lomb-Scargle periodogram of the aforementioned datasets is shown in Fig.~\ref{fig:SigniPerWave} where we see numerous peaks and it seems difficult to find any significant peak against its surrounding. The horizontal lines show the false alarm probability (FAP) of 0.01. The FAP stands for the significance of peak or how rarely the peak could happen from the pure white-noise signal (null hypothesis). The more significant peak has the less FAP. However, this method has a problem as authors indicated: a weak physical cycle (of low amplitude) might be discarded and a strong spurious cycle be elected. \citet{Cameron2019} showed that significant spurious peaks of the Gleissberg and the Suess/de Vries cycles can happen in a noisy model with no intrinsic periodicities except that of the basic 11/22-yr cycle. We try to analyze a reason. 

Let consider an analog signal
\begin{linenomath}\begin{align}
u(t)&= \int^{\infty}_{-\infty} U(\omega) \exp(i \omega t) d \omega,
\end{align}\end{linenomath}
where $U(\omega)$ is the amplitude of the $\omega$ mode. If $u(t)$ is defined in infinite domain of $t\in(-\infty,+\infty)$, $U(\omega)$ could be given by the Fourier transform of $u(t)$ as
\begin{linenomath}\begin{align}
U(\omega) &= \frac{1}{2\pi}\int^{\infty}_{-\infty} u(t)\exp(-i\omega t)d t \label{eq:cft}\\
&=\int^{\infty}_{-\infty} U(\omega')\delta(\omega-\omega')d \omega',\label{eq:delta}
\end{align}\end{linenomath}
where we use $\int^{\infty}_{-\infty} \exp(i\omega t)d t=2\pi \delta(\omega)$ and $\delta(\omega)$ is the Dirac delta function. We can expect that the continuous Fourier transform Eq.~\eqref{eq:cft} gives a spike-like spectrum for a periodic signal. The peak height should be infinitive.

In practice, however, we deal with a length-limited discrete signal. Then $\delta(\omega-\omega')$ in Eq.~\eqref{eq:delta} is replaced with $W(\omega-\omega')$, where
\begin{linenomath}\begin{align}\label{eq:W}
W(\omega)&= \frac{\sin(\omega N/2)}{\sin(\omega/2)}e^{-i \omega(N-1)/2},
\end{align}\end{linenomath}
where $N$ is the length of the time series. Note that we take its absolute value $\vert W\vert$ into account in spectrum and the shape of a peak in the discrete Fourier spectrum will be the same as in Fig.~\ref{fig:Peakform}. For a mono-cyclic time series of $N$ length and unit amplitude, the height of the peak is $N$ and the width of the main lobe is $4\pi/N$ \citep{Orfanidis2010}. Around the main lobe appear even the side lobes numerously, each of them is $2\pi/N$ wide. Therefore we could see innumerable but meaningless peaks in spectrum. (These side lobes can be seen more frequently for non-stationary cycles.) If the time series gets longer, the peak of cycle gets higher and narrower, and we will see a greater number of negligible peaks in spectrum.  

Thus we can claim that the general strength-based spectral analysis, especially the Fourier approach has two drawbacks: first, numerous meaningless small peaks should appear in spectrum and, secondly, the cycles of weak amplitude could be neglected.

Now let consider a typical strength-based spectral analysis - the Lomb-Scargle periodogram. The FAP in the Lomb-Scargle periodogram is deduced under the null hypothesis that a peak could be formed only by a white-noise signal. Here we meet another problem. The spectra of solar activity usually look like AR(1) spectrum, which means an autocorrelation of the time series \citep{Torrence1998}. Thus the shorter periodicities appear weaker and the longer ones appear stronger. However, the FAP are measured equally for the shorter and longer cycles. Furthermore, the smoothing that workers usually perform by moving averages \citep[for example, see][]{Petrovay2010b,Wu2018} lowers the high-frequency (short-period) cycles and raises the low-frequency (long-period) even more (Fig.~\ref{fig:GfcompLSOut}). Then the low frequency could be selected by the FAP more probably than the high frequency. Of course, the noise is usually considered as of high frequency so it seems deserved to neglect them. However, we should remember that very short cycles are existing in solar phenomena such as the 27-day periodicity and even the 3-min oscillation in photosphere. In this sense, we can see that a cycle (e.g. the 5.5-yr or 40-yr periodicities in Fig.~\ref{fig:SigniPerWave}) are significant in a shorter-scale (e.g. the daily) dataset but negligible in a longer-scale (e.g. the yearly or reconstructed) dataset. 

Thus the Lomb-Scargle periodogram seems to have an additional problem: the shorter periodicities in solar activity could be neglected more probably by FAP, and a consistency in FAP of the same cycle between different datasets might be hardly reached.

We need a strength-free spectral analysis and a criterion for significance that is equal for both short and long cycles.

As we have seen above, Eq.~\eqref{eq:cft} and ~\eqref{eq:delta} inspire us that the Fourier transform of signal or the peak height in spectrum can be decomposed into two factors: ``intensity'' and ``coherence'' of the periodic signal. The former stands for the magnitude of signal $u(t)$ or the amplitude $U(\omega)$ and the latter 
\footnote{The stability in phase of a signal of constant period was considered by other workers, for example, the medium-term coherence of stellar activity signals has been addressed by \cite{Gregory2016}.}
for $\exp(-i\omega t)$, $\delta(\omega-\omega')$ or $W(\omega-\omega')$. The $\exp(-i\omega t)$ in Eq.~\eqref{eq:cft} seems to extract the phase of $u(t)$ by multiplication which is accumulated via integral and express how periodically the cycle appears. We can suppose that the periodicity is reflected only on the ``coherence.'' 

As shown in Fig.~\ref{fig:Peakform}, a cycle with unit amplitude makes a peak with the height proportional to the length of time series $N$ (more exactly, proportional to $\sqrt{N}$ by normalization). Of course in that case the cycle is assumed to exist all over the time of signal. This implies that the height of peak in the Fourier approach could be a multiplication of the amplitude and the time length that the cycle persists. Namely, this implies that in a strength-free analysis the height of peak could depend only on the time length that the cycle exists. In fact, \cite{Chol-jun2020a} showed that a temporary or intermittent, i.e. part-time cycle can make a peak in spectrum whose height is in proportion to (square root of) $N/N'$ where $N$ stands for the time length when the cycle exists and $N'$ for the whole period of a zero-padded time series which is composed of the periodic signal followed by zero-signal. In this case, $N/N'$ can be regarded as a proxy for ``coherence'' clearly. Therefore, if we measure the duration that the cycle exists or the fractional duration $N/N'$, we can build a strength-free analysis which excludes the strength of cycle such as the amplitude or the power at all. In the way that we have ever pursued to find any stability of cycle against processing signal \cite[for example, see][]{Chol-jun2020a}, the fractional duration of cycle can refer to as a time stability which expresses a stability of cycle against time progress  

Now we study the time stability of periodicities in solar activity. 
 
\section{The time stability of solar periodicities}

\begin{figure*}
\centering
\subfigure[]{\label{fig:RidgeDSSA27} \includegraphics[width=0.32\textwidth]{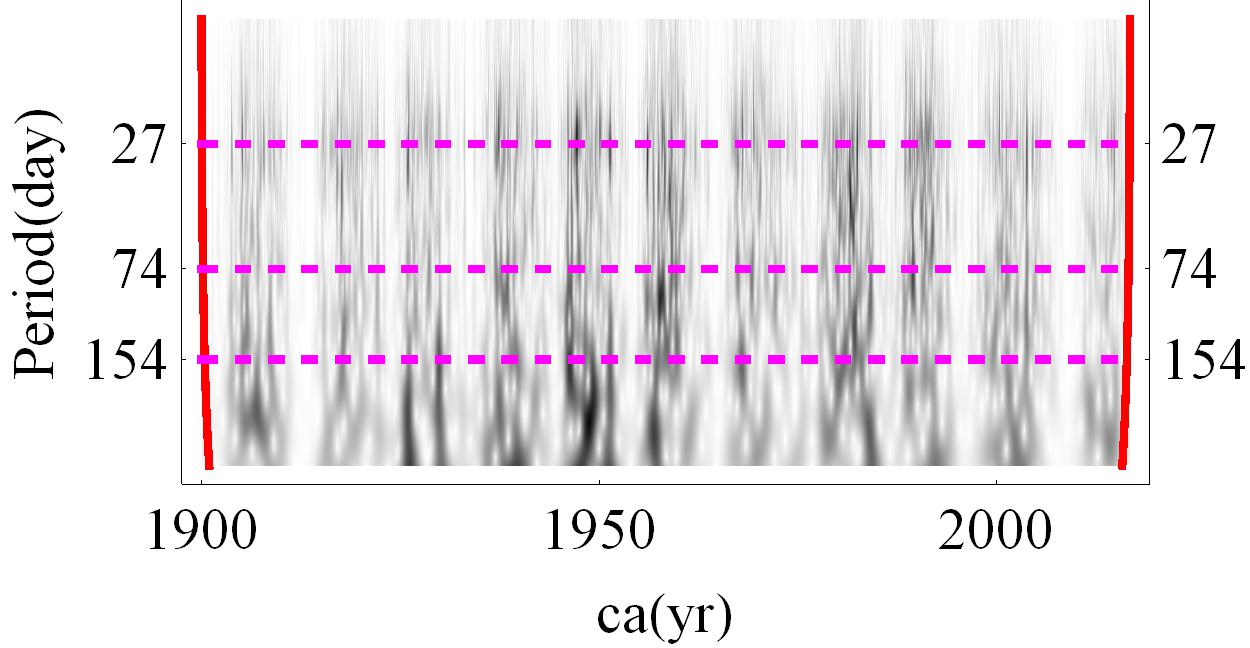}}
\raisebox{-3.0ex}[0ex][0ex]{\subfigure[]{\label{fig:RidgeDGSN1_3} \includegraphics[width=0.33\textwidth]{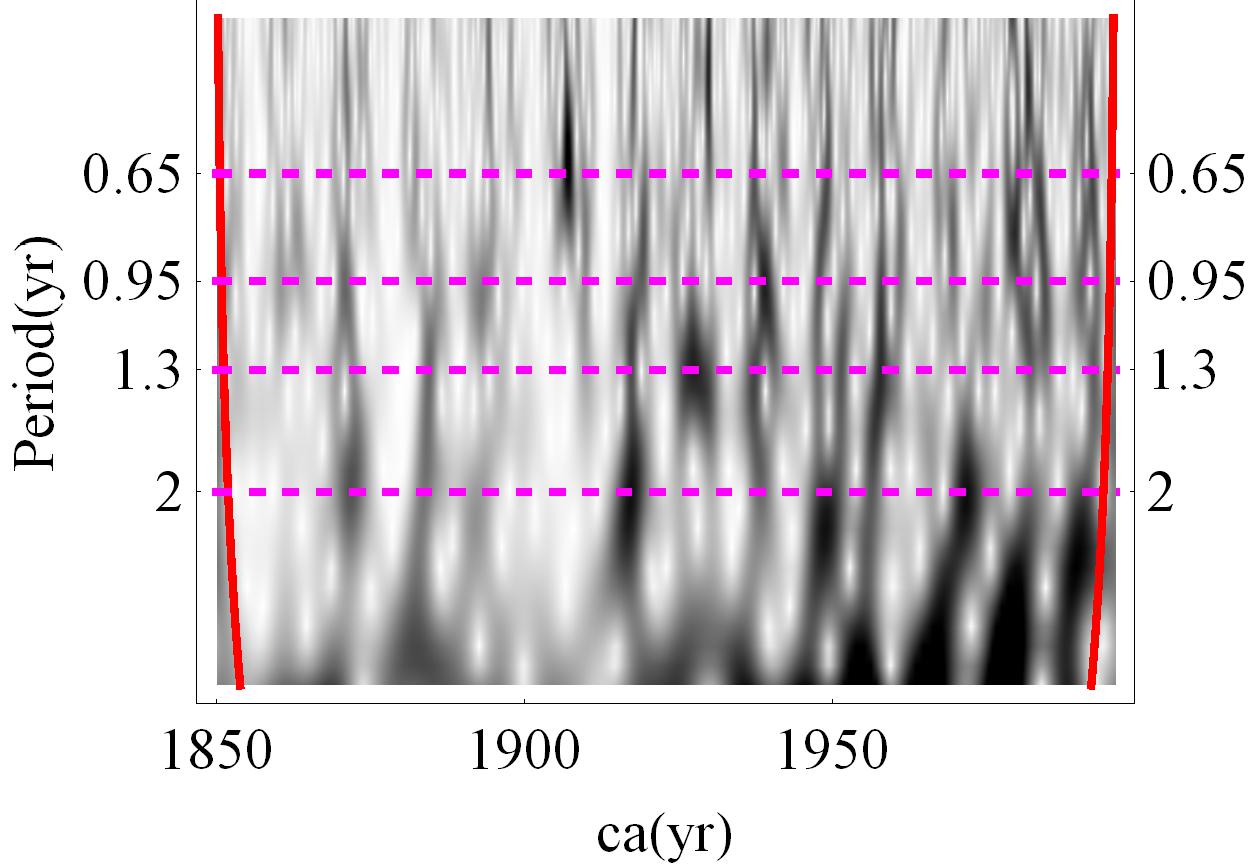}}}
\subfigure[]{\label{fig:RidgeMGSN35} \includegraphics[width=0.32\textwidth]{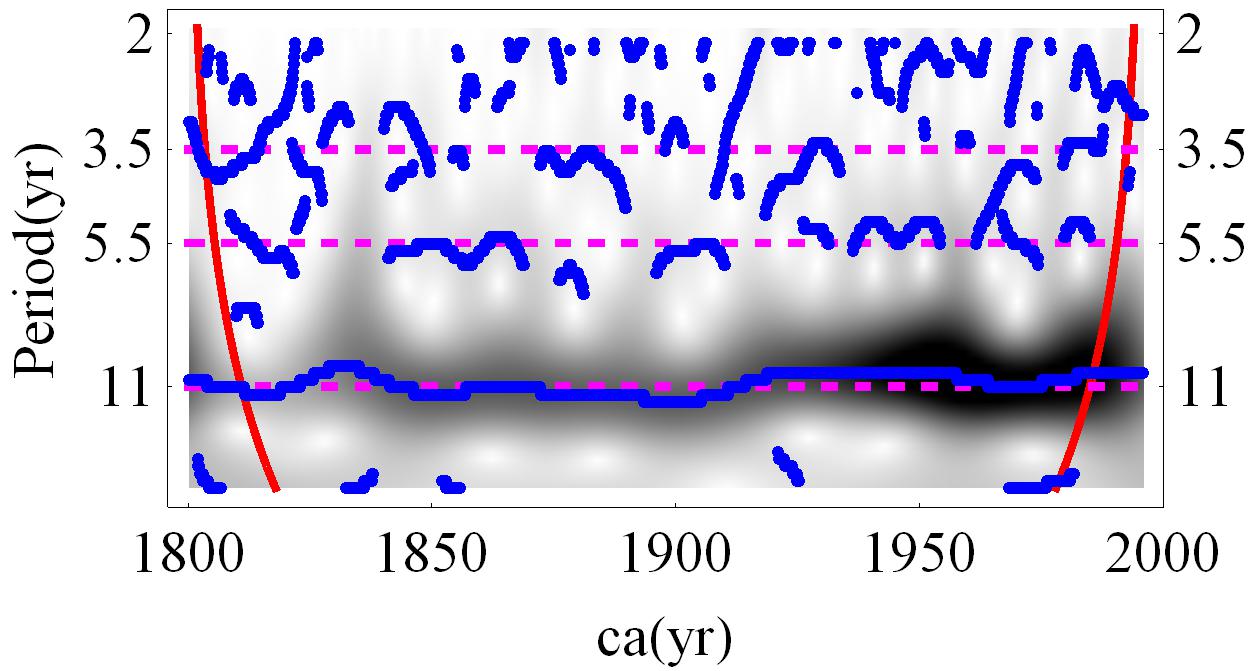}}
\subfigure[]{\label{fig:RidgeMSSN22} \includegraphics[width=0.31\textwidth]{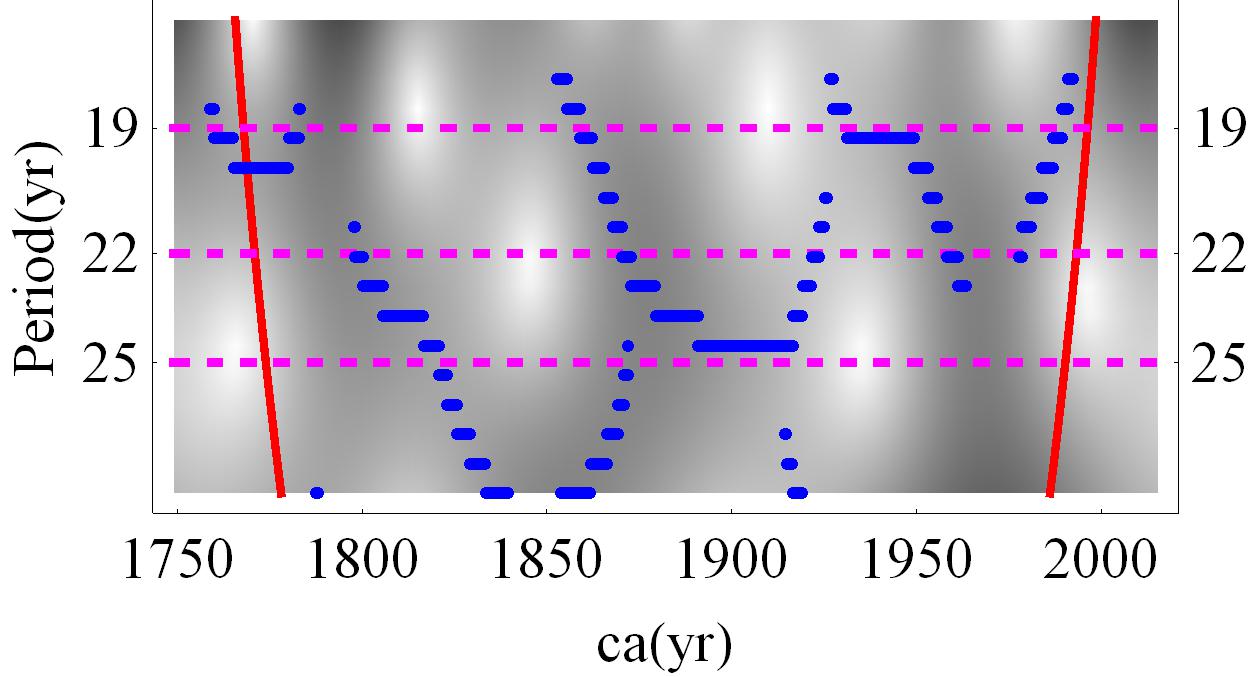}}
\subfigure[]{\label{fig:RidgeYGSN55} \includegraphics[width=0.31\textwidth]{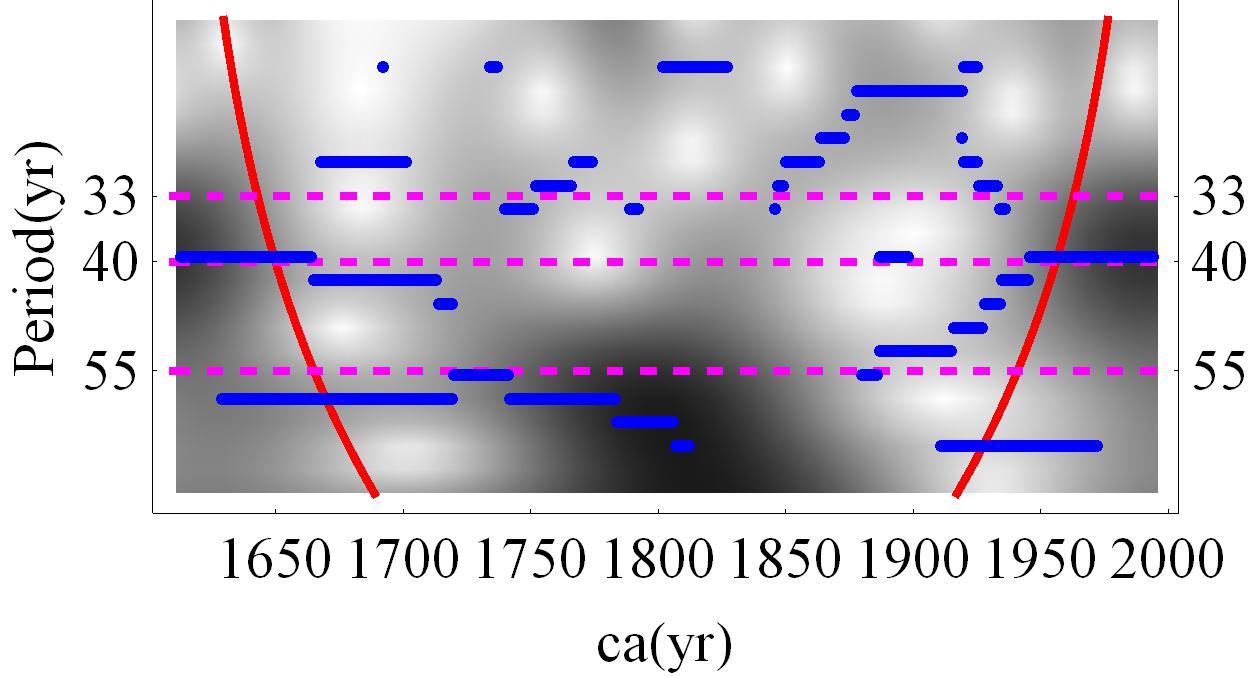}}
\subfigure[]{\label{fig:RidgeRSSN210} \includegraphics[width=0.33\textwidth]{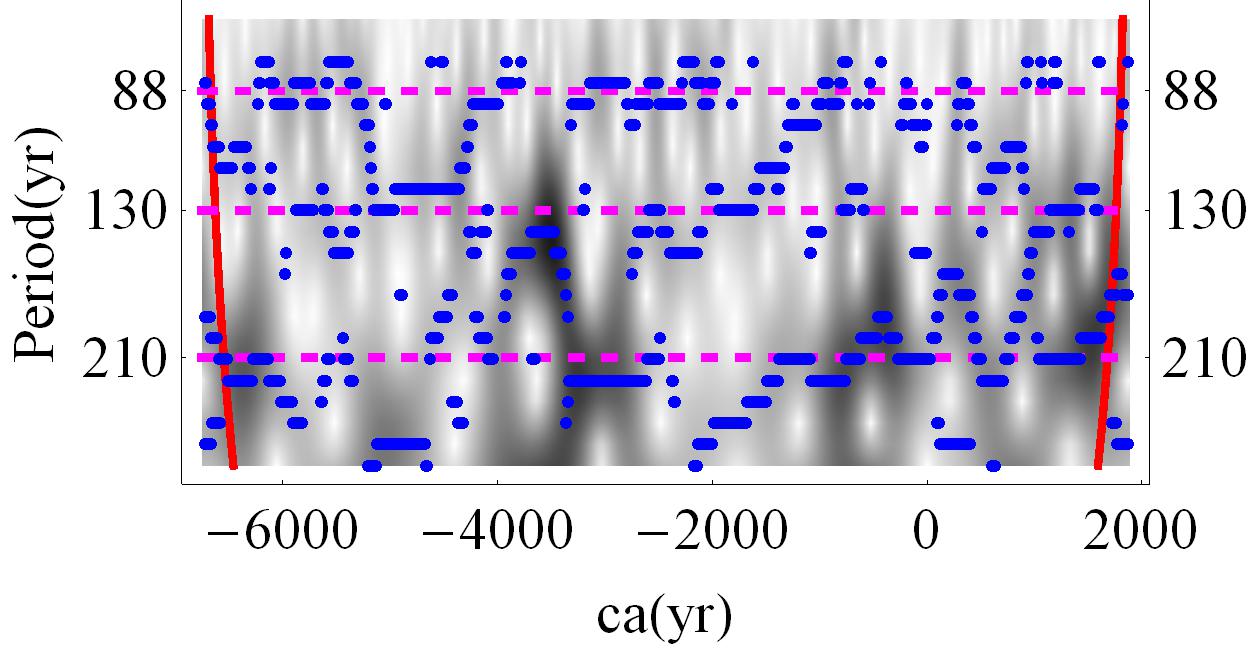}}
\subfigure[]{\label{fig:RidgeRSSN270} \includegraphics[width=0.31\textwidth]{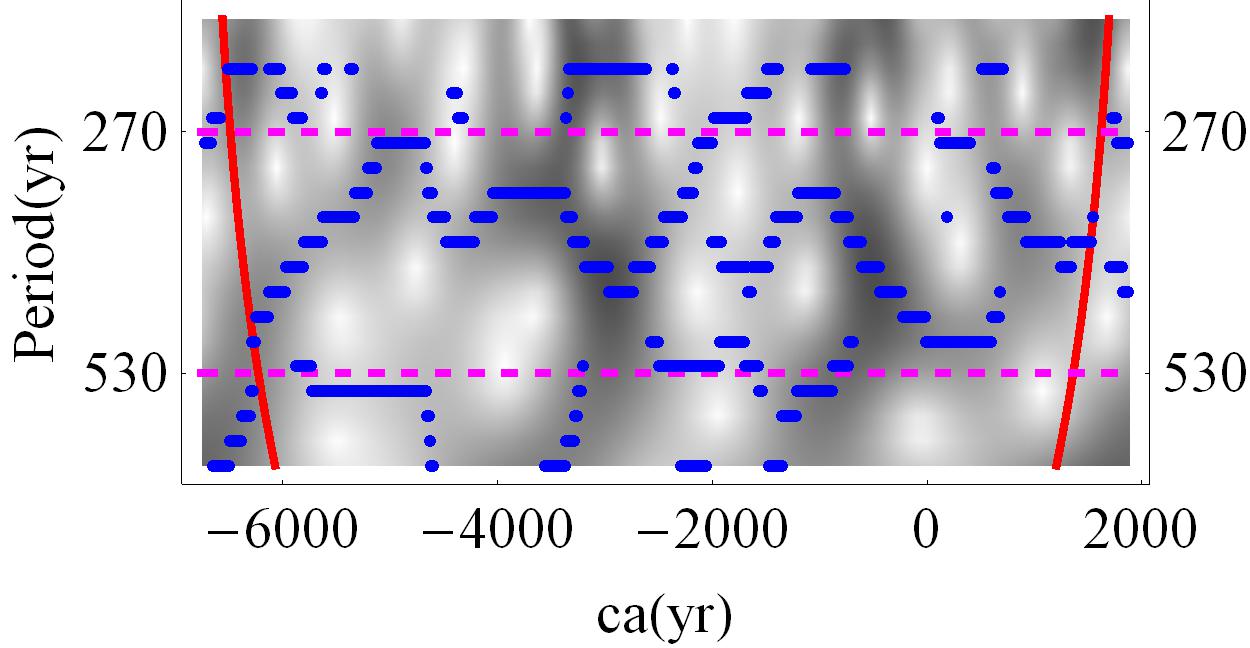}}
\subfigure[]{\label{fig:RidgeRTSI1000} \includegraphics[width=0.31\textwidth]{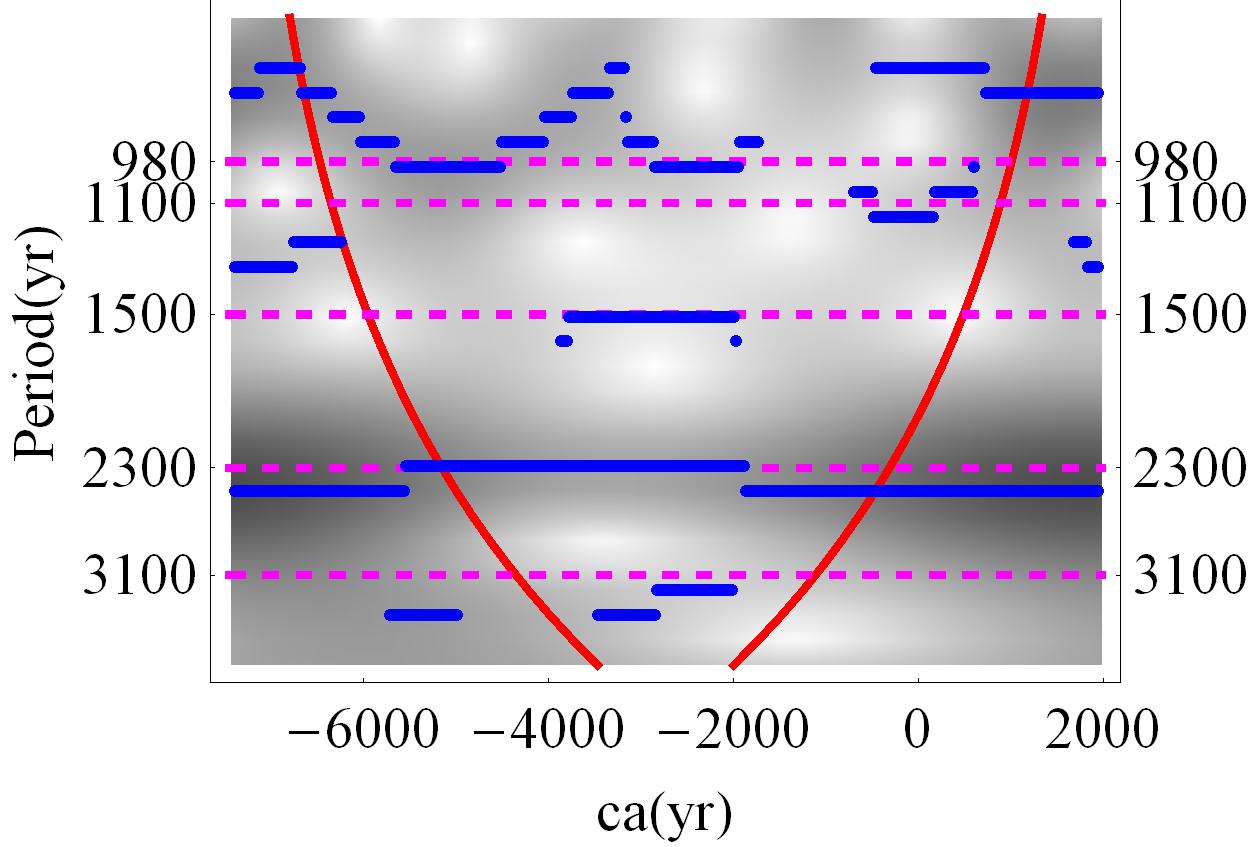}}
 \caption{ \label{fig:RidgeDatasets} Expected solar periodicities and their peak points in the wavelet scalogram. Horizontal dashed magenta lines stand for periodicities and blue dots for peak points which in turn form the ridge. (a) 27, 74 and 154-day periodicities in the DSSA. (b) 0.65, 0.95, 1.3 and 2-yr periodicities in the DGSN. (c) 3.5, 5.5 and 11-yr periodicities in the MGSN. (d) 19 and 25-yr periodicities around 22 yr in the MSSN. (e) 33, 40 and 55-yr periodicities in the YGSN. (f) 88, 130 and 210-yr periodicities in the RSSN. (g) 270 and 530-yr periodicities in the RSSN. (h) 980, 1100, 1500, 2300 and 3100-yr periodicities in the RTSI. The background density plots stand for the wavelet scalogram for the datasets. The red curves delimit the COI. For (a) DSSA and (b) DGSN, denoted are only the periodicities and no peak points because of lack of computing capability. The number of voices per octave is set to 20 in (c) and (d) for enhancing resolution to discern the 19 and 25-yr periodicities but 10 in other cases.}
\end{figure*}

To study the time stability of cycle, we use the time-frequency analysis, e.g. the wavelet analysis. In the wavelet scalogram we can define a ridge, which is formed from the peaks corresponding to the part-time cycles. Figure~\ref{fig:RidgeDatasets} shows some already-reported periodicities and their ridges in the wavelet scalogram of the aforementioned datsets.

As we said above, the fractional duration of cycle can be defined as a time stability (shortly, TS) of cycle
\begin{align}\label{eq:ac}
\text{the time stability of cycle} = \frac{\text{the accumulated duration of cycle}}{\text{the whole duration of signal}} (= \text{the fractional duration of cycle}) ,
\end{align}
The time stability can measure how long the cycle persists during the time of signal. If a cycle exists absolutely ``full-time,'' its time stability should be unity. 
The occurrence of cycle has no relation with the intensity of cycle so that we could perform a strength-free analysis.

In many softwares such as \texttt{Mathematica} or \texttt{MATLAB}, the continuous wavelet scalogram consists of pixels with $\Delta t$ and $\Delta s$, where $t$ stands for the time and $s$ for the scale or the period of cycle. We can obtain the numerator by counting the number of pixels on ridge for the cycle. In this sense, the denominator should be the length of the time series. Here we could neglect the effect of COI (the cone of influence) because the zero-padding within it does not affect the period of cycle. 
We set the number of voices per octave in wavelet analysis to 10 (Fig.~\ref{fig:RidgeDatasets}). Then the period of cycle is determined to extent of $\sim1.07$ times.

\begin{figure*}
\centering
\subfigure[]{\label{fig:tstabStatGrphOutRSSN} \includegraphics[width=0.3\textwidth]{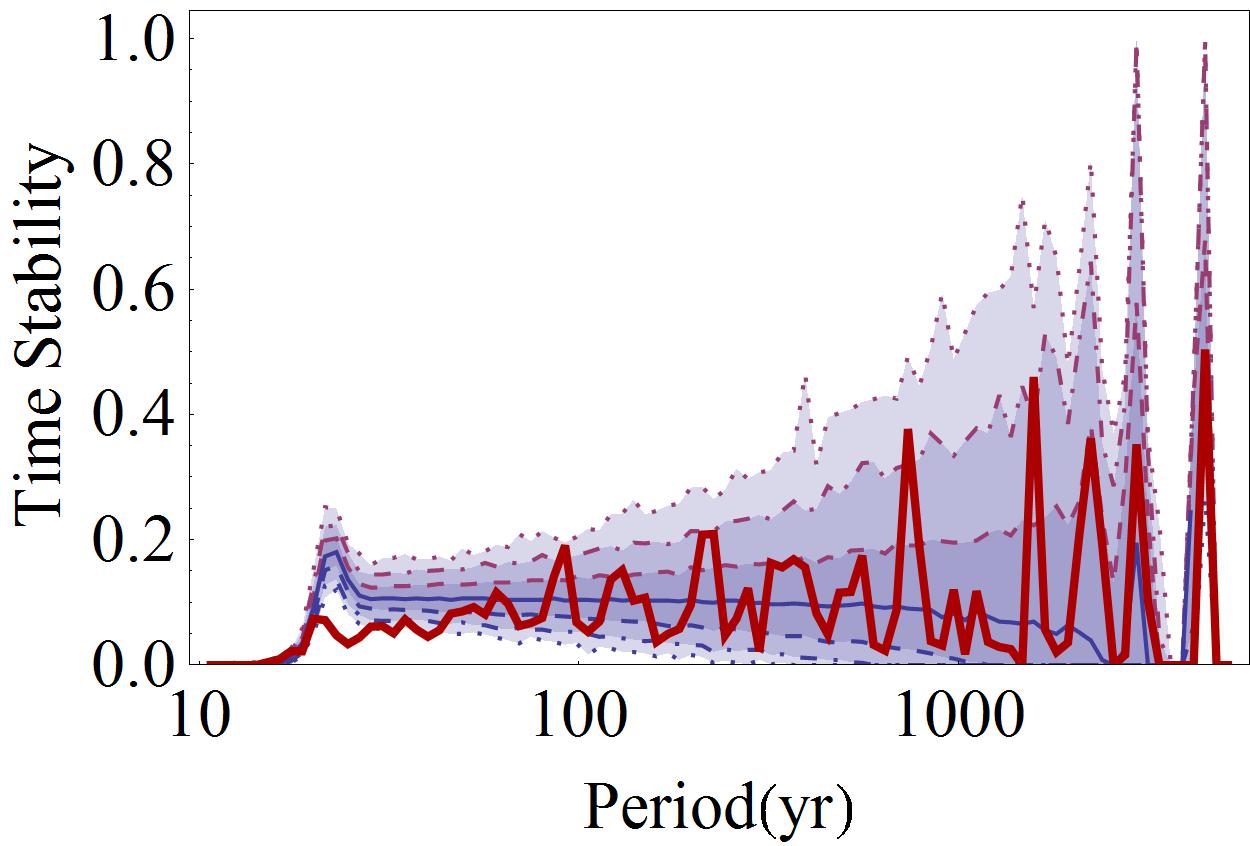}}
\subfigure[]{\label{fig:tstabStatGrphOutRTSI} \includegraphics[width=0.3\textwidth]{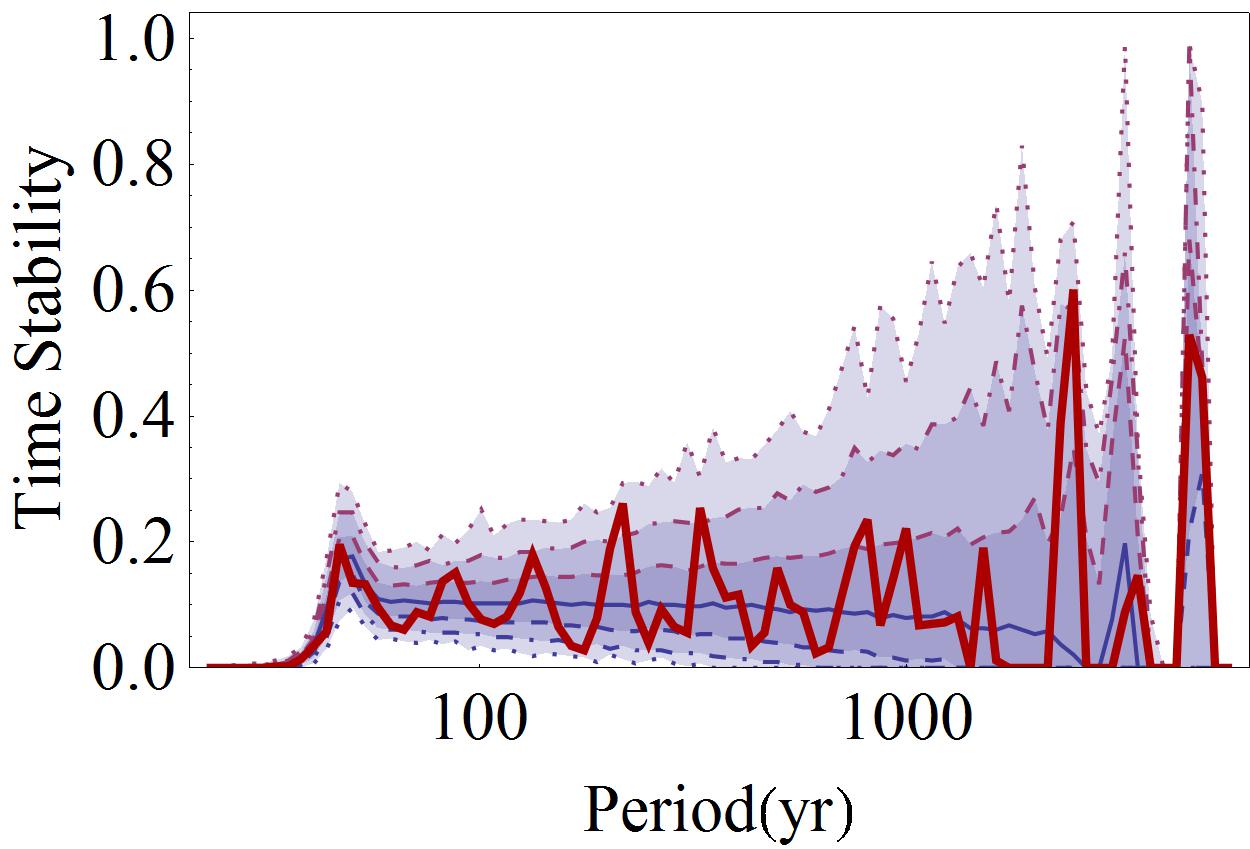}}
\subfigure[]{\label{fig:tstabStatGrphOutYSSN} \includegraphics[width=0.3\textwidth]{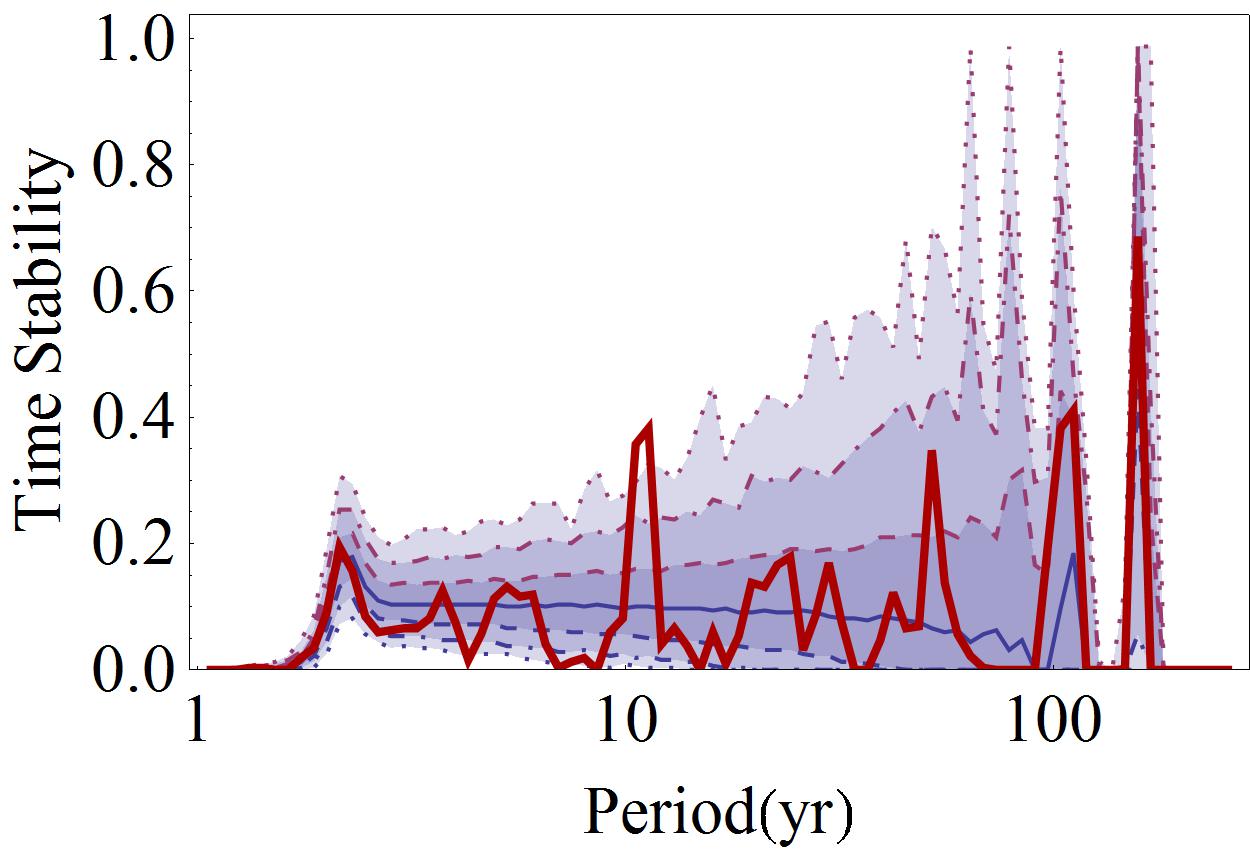}}
\subfigure[]{\label{fig:tstabStatGrphOutYGSN} \includegraphics[width=0.3\textwidth]{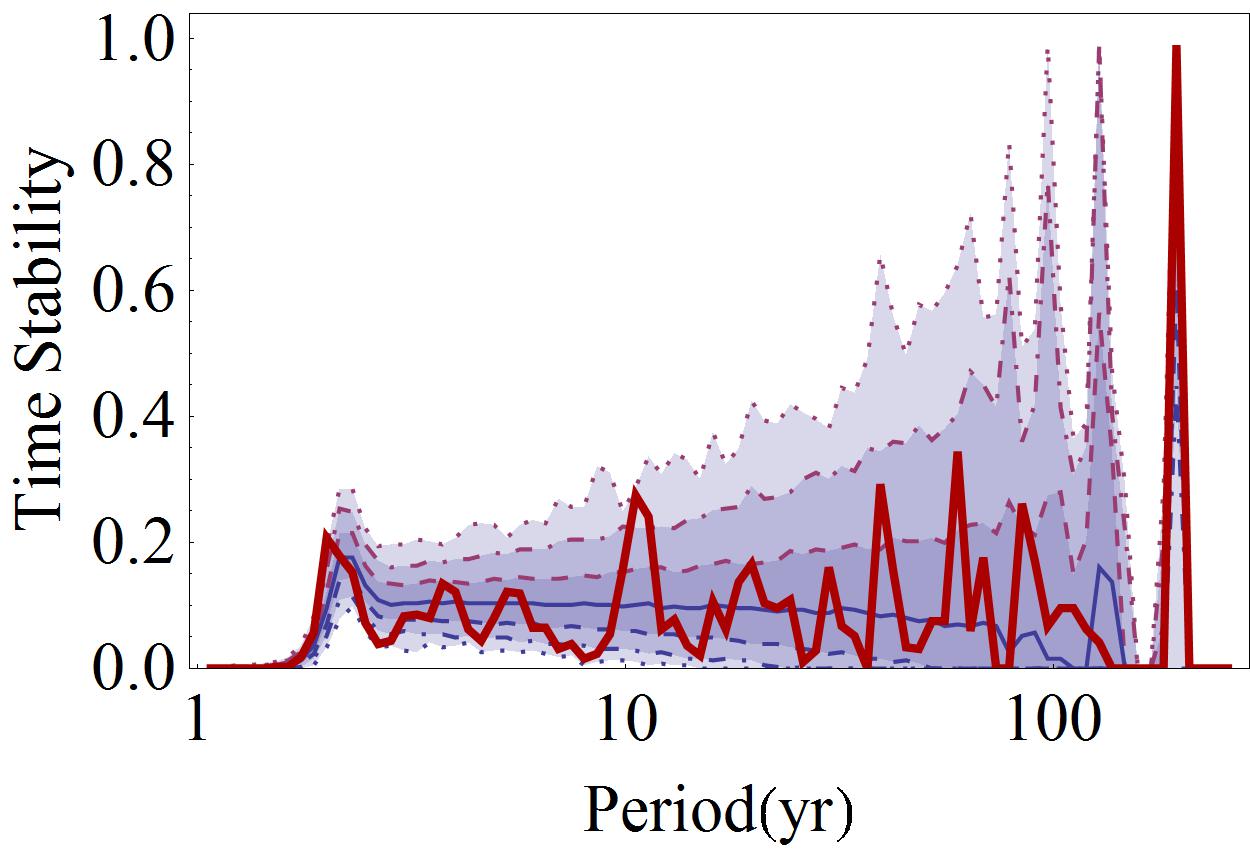}}
\subfigure[]{\label{fig:tstabStatGrphOutMSSN} \includegraphics[width=0.3\textwidth]{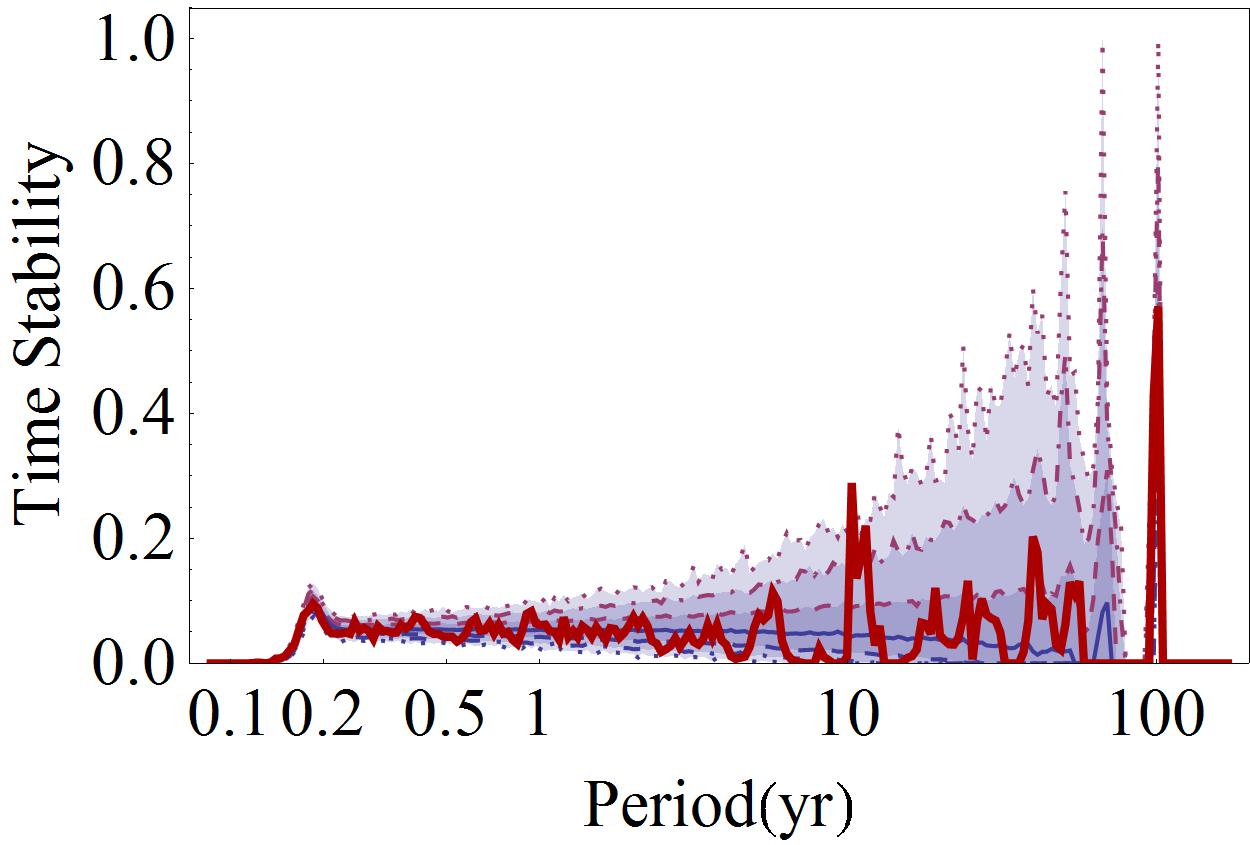}}
\subfigure[]{\label{fig:tstabStatGrphOutMGSN} \includegraphics[width=0.3\textwidth]{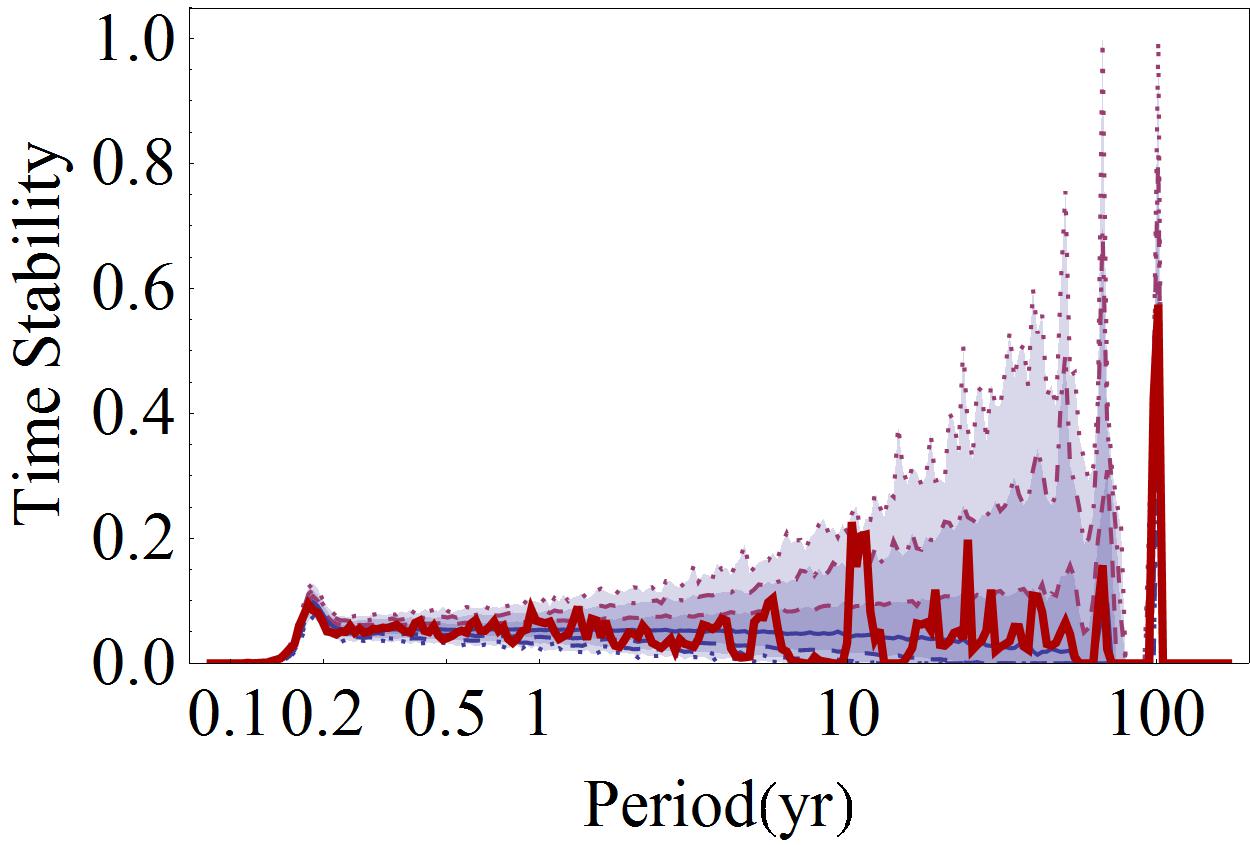}}
\subfigure[]{\label{fig:tstabStatGrphOutDSSN} \includegraphics[width=0.3\textwidth]{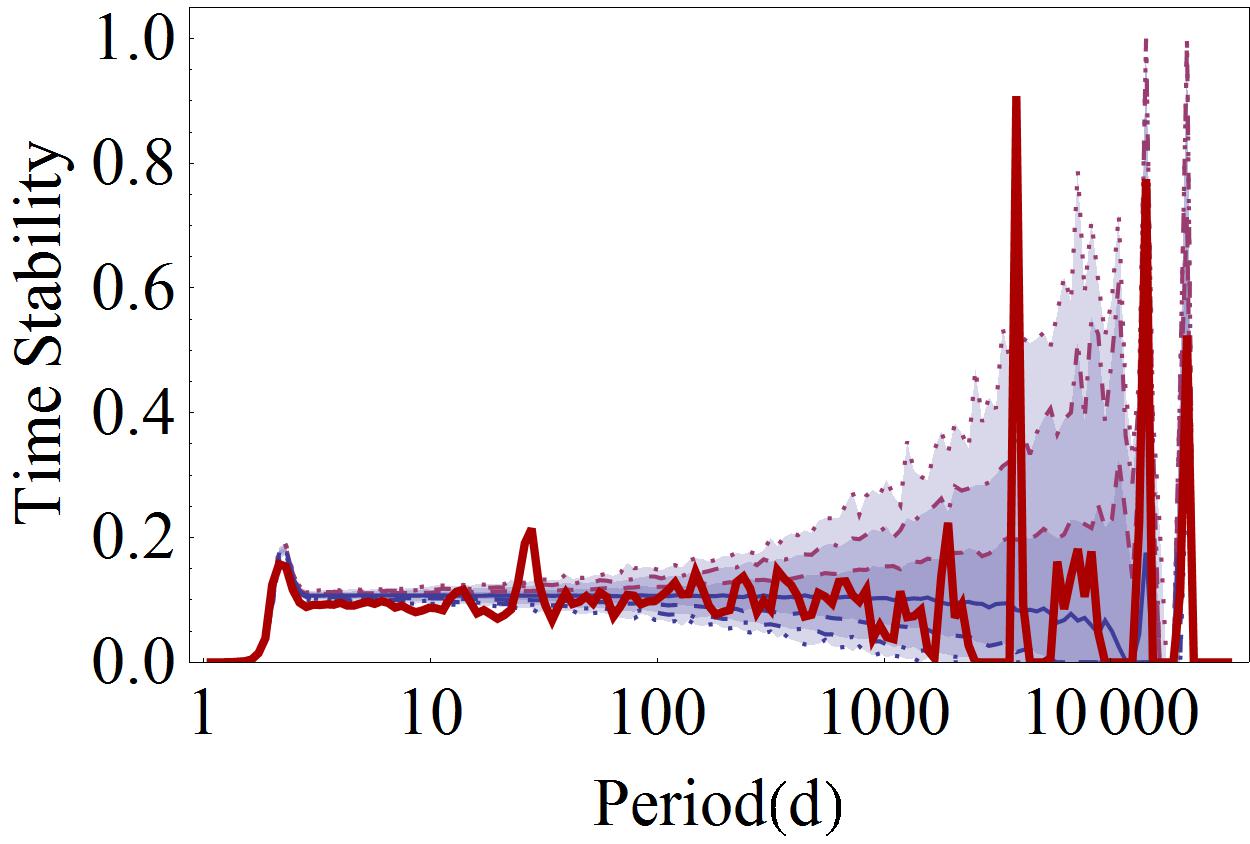}}
\subfigure[]{\label{fig:tstabStatGrphOutDGSN} \includegraphics[width=0.3\textwidth]{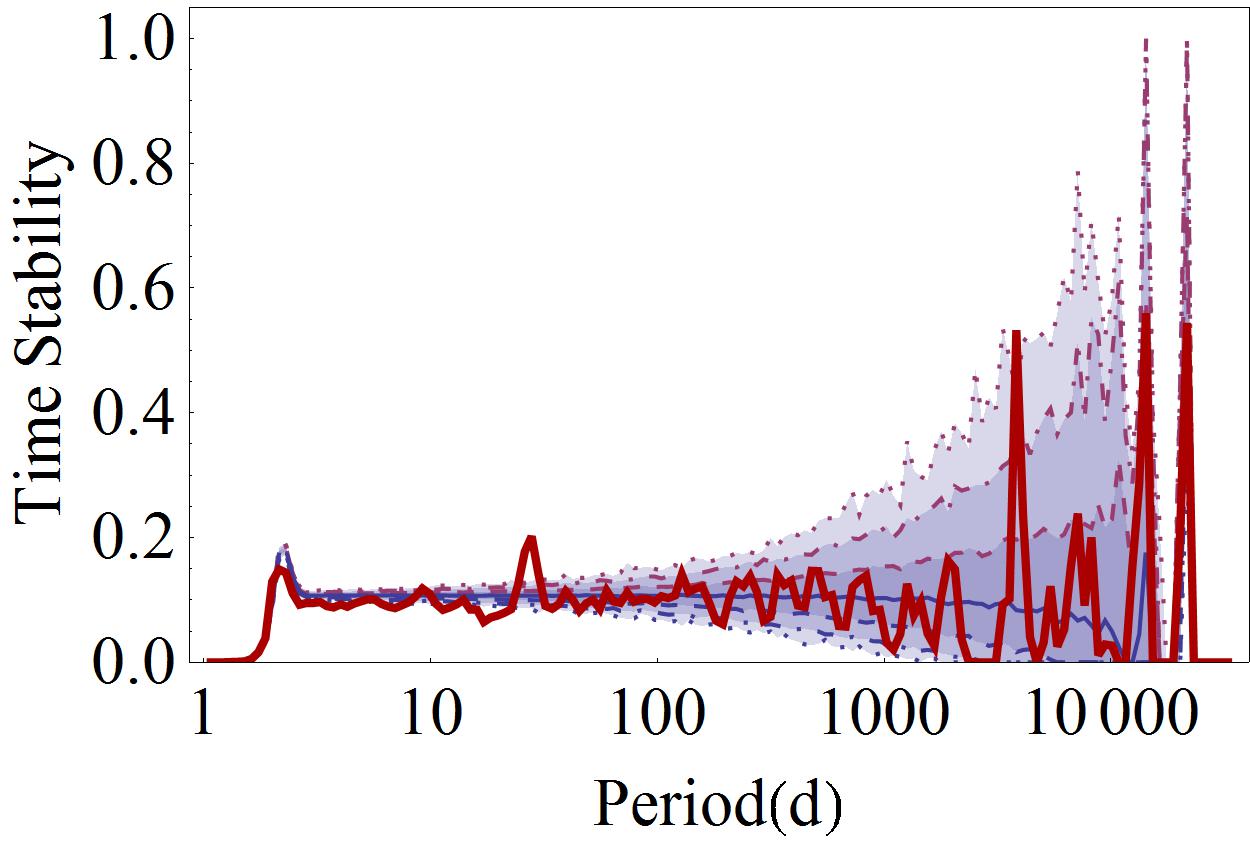}}
\subfigure[]{\label{fig:tstabStatGrphOutDSSA} \includegraphics[width=0.3\textwidth]{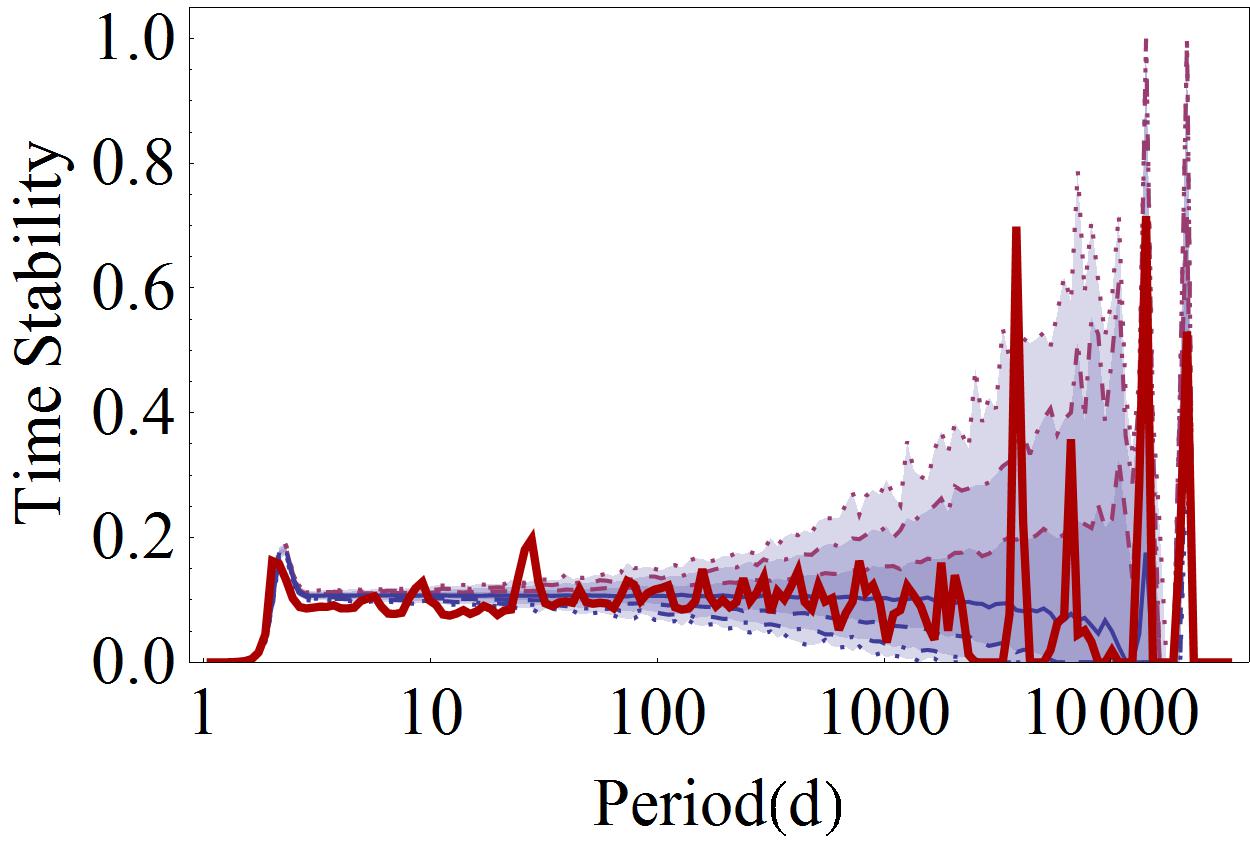}}
 \caption{ \label{fig:tstabStatGrphOut} The time stability of solar periodicities with the confidence levels in the different datasets. (a) RSSN, (b) RTSI, (c) YSSN, (d) YGSN, (e) MSSN, (f) MGSN, (g) DSSN, (h) DGSN, (i) DSSA. The red curve stands for the time stability in the given dataset and the blue curve for the most probable (mode) time stability. The shaded contours stand for $\sigma, 2\sigma$ and $3\sigma$ C.L.  }
\end{figure*}

We inspect statistics of the time stability. The null hypothesis is that the given time stability can be reproduced in pure white-noise time series. Of course, in this case the higher time stability could be obtained more rarely. We perform a Monte-Carlo (MC) simulation with 1000 white-noise time series with the same length and variance to each dataset, respectively. In Fig.~\ref{fig:tstabStatGrphOut} the time stability is shown with the confidence levels (C.L.). We can find some properties of the diagrams.

First, the time stability and its C.L. do not have relation with the amplitude of time series. It coincides with the motivation of the time stability. However, the length of time series affects the time stability: the longer time series would be, the range of C.L. should become narrower.

Secondly, the time stability has a trend to raise from 0 to 1 according to the period of cycle. This trend is similar to the AR(1) spectrum: the short cycle has low time stability while the long cycle has high time stability. \footnote{This could be caused from the wavelet analysis because the random time series used in the simulation have nearly vanished autocorrelation.} This could explain why long spurious cycles appear significantly more often than short ones in most spectral analysis.

Thirdly, however, the AR(1)-similar trend is not all. 
We can find several jumps in the diagrams: the subharmonic of the time step of time series (which is 2 times as long as the time step) and the harmonics of the whole period of time series (which are a half, a third and so on of the whole period of time series) are obviously shown. This gives a possibility to distinguish the spurious cycles related to the structure of time series itself. For example, 20-yr for the RSSN and 44-yr for the RTSI are the subharmonics of the time step in each dataset. Those cycles are related not to the solar activity, but only to the sampling process so we can neglect them. 

\begin{figure*}
\centering
\subfigure[]{\label{fig:GfcompTSOut} \includegraphics[width=0.3\textwidth]{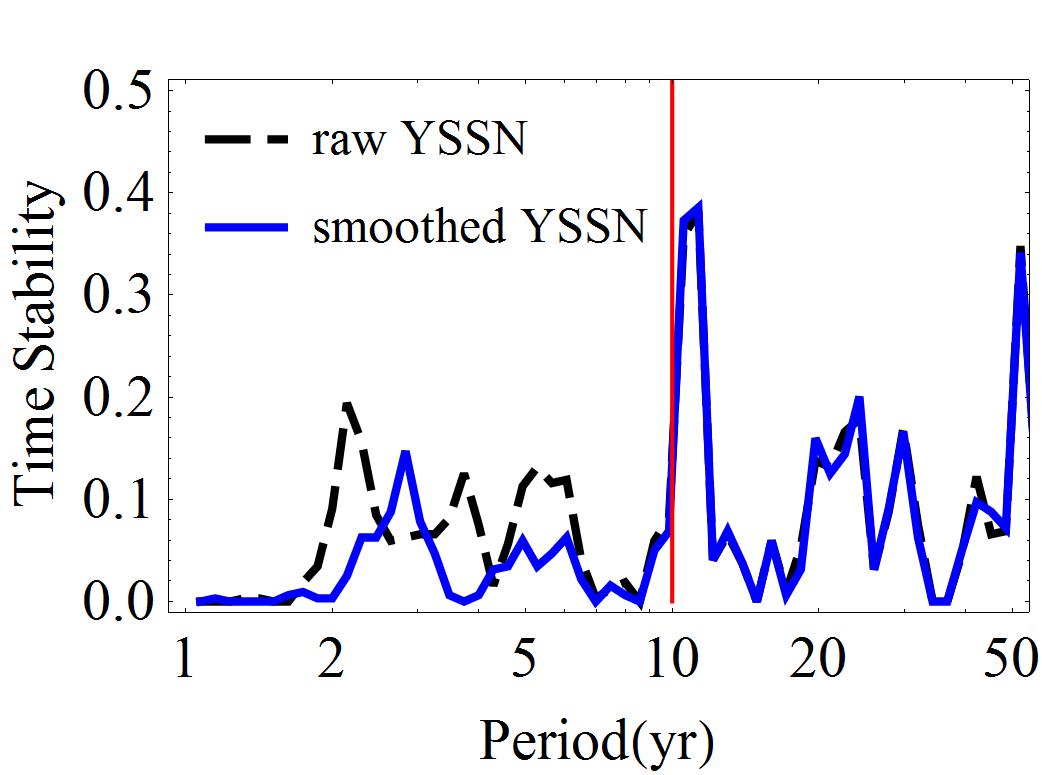}}
\subfigure[]{\label{fig:GfcompLSOut} \includegraphics[width=0.3\textwidth]{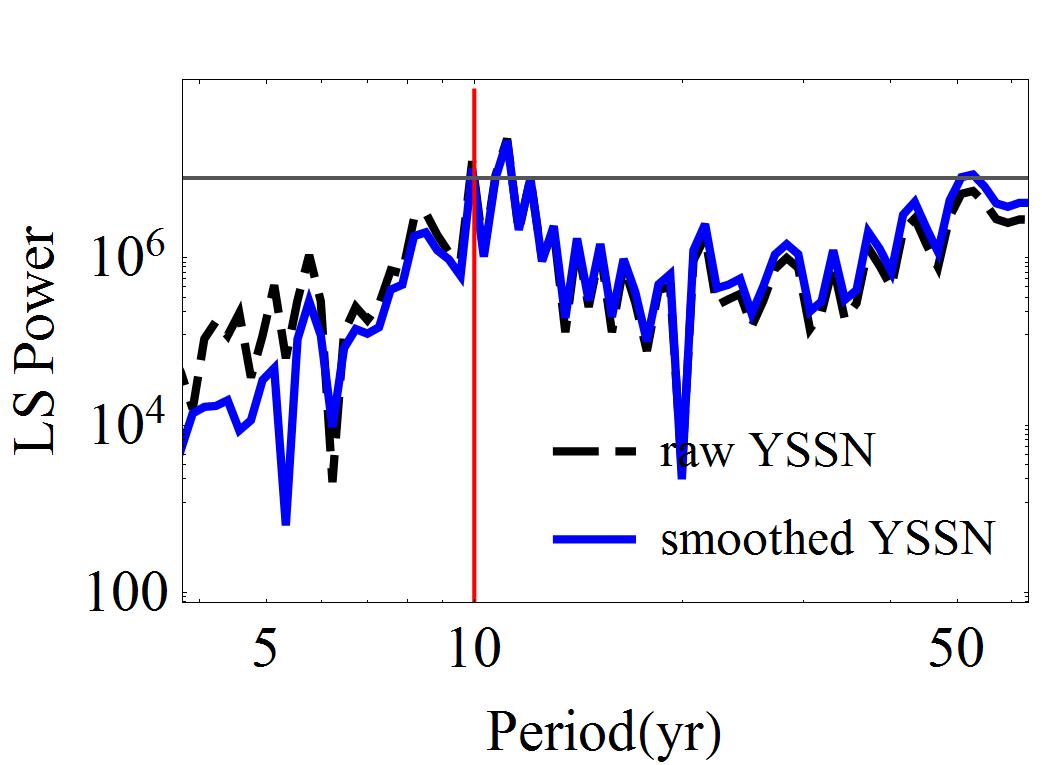}}
\subfigure[]{\label{fig:tstabStatGrphOutYSSN20} \includegraphics[width=0.3\textwidth]{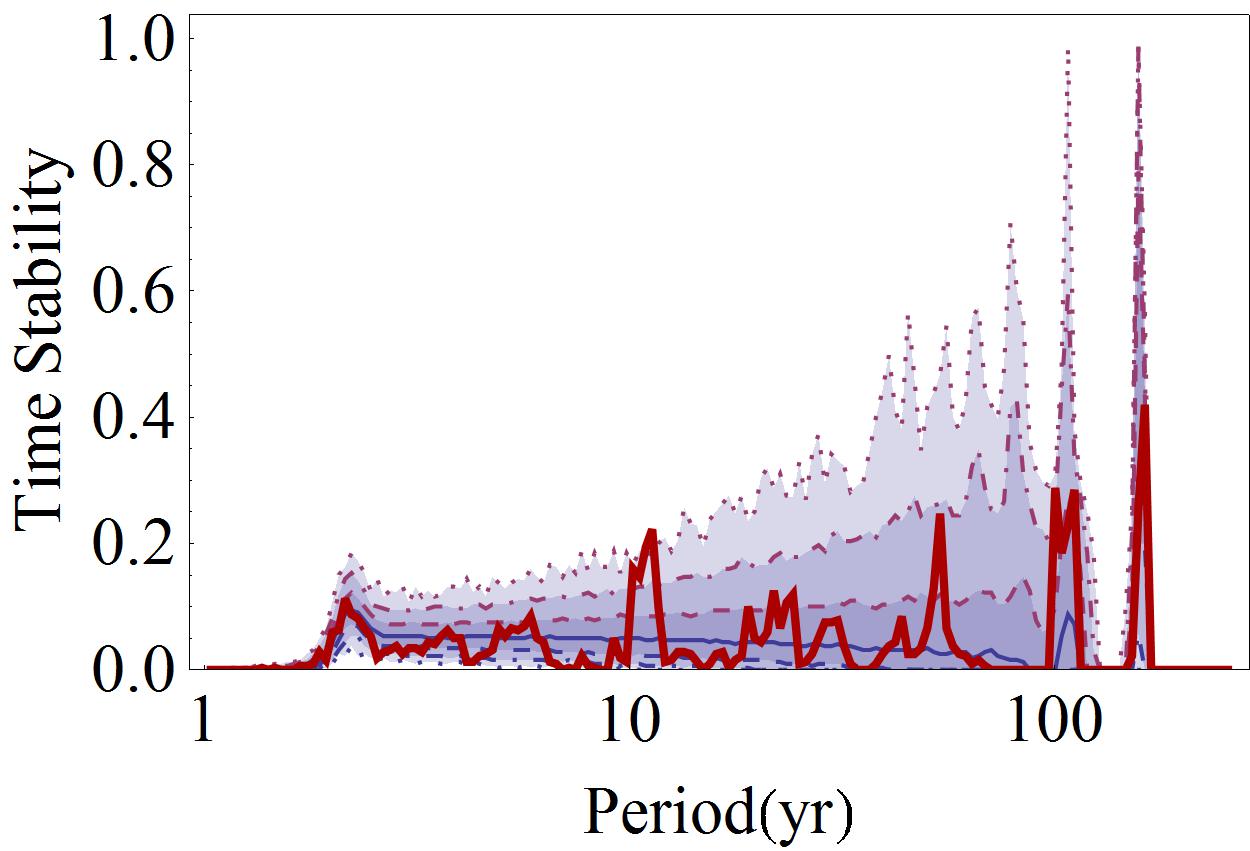}}
 \caption{ \label{fig:TSprocess} Signal processing changes the spectra. (a) The time stability of the raw YSSN  dataset (dashed) and smoothed YSSN (solid) by secular smoothing. (b) The Lomb-Scargle periodogram for the both datasets. The smoothing lowers the short cycle and raises the long cycle. The horizontal line stands for FAP of 0.01. In (a) and (b) the red vertical line implies 10 yr as a potential pivotal cycle. (c) If we raise the number of voices per octave from 10 to 20 in wavelet analysis, the time stability lowers and ranges of C.L. become narrower in comparison with Fig~\ref{fig:tstabStatGrphOutYSSN}.  }
\end{figure*}

Fourthly, the time stability or its range of C.L. can vary in signal processing. For example, we apply a secular smoothing, i.e. the Gleissberg filter \citep{Gleissberg1967} to the YSSN dataset;
\begin{align}\label{eq:Gfilt}
\left\langle R \right\rangle^{(n)}=\frac{1}{8}\left( R^{(n-2)}+2R^{(n-1)}+2R^{(n)}+2R^{(n+1)}+R^{(n+2)}\right).
\end{align}
The time stability for cycles shorter than $\sim 10$ yr undulates while that for longer cycles remains almost intact (Fig~\ref{fig:GfcompTSOut}). In contrast of this, the Lomb-Scargle power lowers for the shorter cycles and raises for the longer cycles after the smoothing (Fig~\ref{fig:GfcompLSOut}) while FAP remains intact. For another example, if we raise the number of voices from 10 to 20 in wavelet analysis, we can see that the overall time stabilitiy lowers and C.L. ranges become narrower (Fig~\ref{fig:tstabStatGrphOutYSSN20}).

Based on the statistics of the time stability, we evaluate stability-based significance of solar periodicities (Fig~\ref{fig:SignifStabPerWave}). Here we can find some properties.
First, the cycles have various significance for the time stability. Secondly, the same cycles seem to have the similar significance even in the different datasets. Finally, the number of peaks seems much smaller than in classical strength-based spectrum such as the Lomb-Scargle periodogram. As we mentioned in Sec.~\ref{sec:risk}, in the strength-based spectrum the width of peak is inversely proportional to the length of time series and innumerable peaks appear in spectrum of random signal. However, in the wavelet analysis the width of peak is determined only by the number of voices per octave.

\begin{figure*}
\centering
\subfigure[]{\label{fig:SignifStabPerWave} \includegraphics[width=0.9\textwidth]{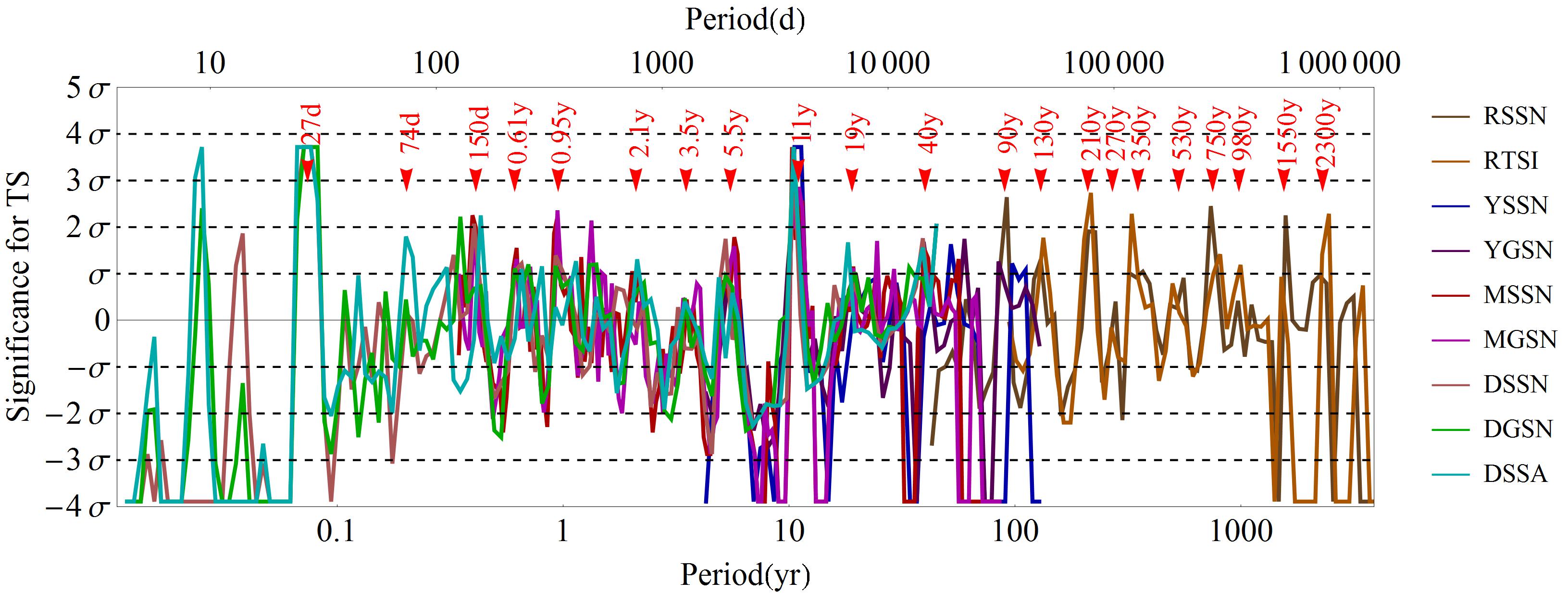}}
 \caption{ \label{fig:SignifStabPerWave} The significance for the time stability of solar periodicities. Cycles common in every datasets are pointed by red arrows. The negative significance stands for minima in spectrum of the time stability while the positive one for maxima.}
\end{figure*}

We can find that most periodicities claimed in the exterior solar activities such as the solar wind in Sec.~\ref{sec:intro} appear also in Fig.~\ref{fig:SignifStabPerWave}. 
The 27-day and 11-yr periodicities are most significant with $\sim4\sigma$ C.L. Though their TS reach much less than unity, we can conclude that those periodicities persist throughout the time series of the datasets. As we know, they are obvious physical periodicities. In fact, the 27 day is just the cycle of average solar rotation.

No other periodicities reach 3$\sigma$ significance, however, appearing commonly in the different datasets:
A 74-day (1$\sigma$), Rieger's 150-day (0.41-yr, 2$\sigma$), a 220-day (0.61-yr, 1$\sigma$), a 345-day (0.95-yr, 2$\sigma$), the quasi-biennial 2.1-yr (1$\sigma$), the 3.5-yr (<1$\sigma$), the 5.5-yr (2$\sigma$), a 19-yr (1$\sigma$), a 40-yr (1$\sigma$), the Gleissberg 90-yr (1$\sigma$), Attolini's 130-yr (1$\sigma$), the Suess/de Vries 210-yr (2$\sigma$), a 270-yr (<1$\sigma$), the 350-yr (1$\sigma$), the 530-yr (1$\sigma$), the 750-yr (<2$\sigma$), the Eddy 980-yr (1$\sigma$) and 1100-yr (<1$\sigma$), a 1550-yr (<2$\sigma$) and the Hallstat 2300-yr (<2$\sigma$) periodicities.
\footnote{A 3100-yr and even the Hallstat cycle of $\sim 2300$ yr might be the second or third harmonics of the whole period of the reconstructed datasets. The 1550-yr periodicity which might be the harmonic of the 3100-yr is more significant than the 3100-yr itself. Thus we neglect the 3100-yr periodicity. }
Most already-claimed periodicities are involved. The analysis shows also new periodicities such as the 74-day, 0.61-yr, 0.95-yr, 40-yr and 1550-yr periodicities.
\footnote{In the RSSN and RTSI datasets there appears a new 1550-yr periodicity, which does not appear in the periodogram (Fig.~\ref{fig:SigniPerWave}). This periodicity might be related to the 1500-yr cycle in the climate change \citep{Bond1997} or remained in the reconstruction process of the past solar activity.} 
Meanwhile, some already reported periodicities have been missed. There are no long-term periodicities found in the reconstructed datasets, though it need more datasets for comparison.

Could those periodicities be simply the noises? Less significance may say yes. However, we can give some periodicities a physical reasoning. For example, the 5.5-yr periodicity is surely the harmonic of the 11-yr periodicity.
\footnote{Instead, the 22-yr cycle, which is the subharmonic of the 11-yr periodicity and appear in most spectral analyses, disappears but meanders between 19 yr and 24 yr (Fig.~\ref{fig:RidgeMSSN22}). Recall that the 19 and 24-yr periodicities correspond to valleys but the 22-yr to a peak in the periodogram (Fig.~\ref{fig:SigniPerWave}). The 22-yr periodicity is probably different from the Hale cycle, because the latter is only a magnetic cycle and can appear as the 11-yr periodicity in spectra.}
Besides, strong annual modulations in the solar polar field strength are due to the tilt of the solar equator to the Ecliptic, which could be a cause of the 0.95-yr periodicity \citep[for example, see][]{Petrovay2010b}.
So we could attribute the less significance of some cycles to a drawback of our approach: the significance in this test can not determine whether the signal is consistent with the noise or not. 

Some solar periodicities might appear in a modulation with other periodicity. For example, the less significant 74-day periodicity seem to appear intermittently with cycle of 11 yr while the 210-yr modulated by the 2300-yr, respectively (Fig.~\ref{fig:RidgeDatasets}). This modulation should shorten the possible total duration of cycle and lower its significance. We might need a more effective approach which could filter all the seemingly remained physical periodicities. However, the periodicities having the different properties, so the detecting ways should be different according to the periodicities.

\section{Conclusion and discussion}
Traditional spectral analysis can be classified as strength-based where the strength means power or amplitude. If a weak cycle is included in signal and even if the signal is random, we can never find a negligible peak corresponding to this cycle embedded in the background of innumerable noisy peaks. In order to find weak cycles, we have to have a spectral analysis to exclude the amplitude information of cycles and evaluate only their periodicity. 

In this paper we proposed a time stability of cycle as a criterion to evaluate the periodicity of cycle. 
The time stability implies a fractional duration or stationarity of intermittent or part-time cycle. 

The time stability analysis have some advantages. First, the spectrum of time stability is simple and obvious. While there are innumerable peaks in strength-based spectrum, the peaks in the time-stability spectrum have moderate width enough to be distinguish and the spectrum seems simpler. The second advantage is an ability to distinguish spurious cycles. In fact, many structural harmonics and subharmonics are revealed so to remove them easily. Thirdly, the statistics of the time stability could give a significance consistent in different datasets and equal for shorter and longer cycles.

We inspect the sunspot-related datasets and found several time-stable periodicities. The most significant and enough significant periodicities are only the Dicke 27-day (4$\sigma$) and the Schwabe 11-yr (4$\sigma$) cycles. In spite of their different amplitude, those cycles show the similar significance for the time stability. Most periodicities claimed in the exterior solar activities appear commonly in the different datasets of the interior solar activity with less significance. This shows that those periodicities could be regarded as the realization noises, and those periodicities or noises are originated from the interior solar activity.

However, some meaningful periodicities also have less significance. It is a drawback that this approach can not give a reasoning to it. Here we did not give an analytic interpretation for variation of the time stability according to the period of cycle in various condition. A statistics of consistency between the different datasets also remained open. 

Beside the time stability, we could find another stability of cycle. We expect a more effective way for spectral analysis of stochastic solar activity.

\section*{Acknowledgements}
\addcontentsline{toc}{section}{Acknowledgements}

K. Chol-jun has been supported by \textbf{Kim Il Sung} University during the investigation. We are grateful to the journal's editors and anonymous reviewers whose indications 
were very helpful to lead us to complete the idea step by step.

\section*{Data availability}
Data used in this paper are available at the website addresses indicated or by corresponding with the authors.

\end{document}